\documentclass[%
 reprint,
 longbibliography,
 amsmath,amssymb,
 aps,
]{revtex4-1}

\usepackage{xcolor}
\usepackage{graphicx}% Include figure files
\usepackage{dcolumn}% Align table columns on decimal point
\usepackage{bm}% bold math
\newcommand{\E}{\varepsilon}

\newcommand{\B}{\beta}

\renewcommand{\bf}{\textbf}

\begin{document}

\preprint{APS/123-QED}

\title{Designing Metagratings Via Local Periodic Approximation: \\
From Microwaves to Infrared}% Force line breaks with \\
%\thanks{A footnote to the article title}%

	\author{Vladislav Popov}
	\email{uladzislau.papou@centralesupelec.fr}
	\affiliation{%
		SONDRA, CentraleSup\'elec, Universit\'e Paris-Saclay, F-91190, Gif-sur-Yvette, France
	}%
	\author{Marina Yakovleva}
%	\email{marina.yakovleva@c2n.upsaclay.fr}
	\affiliation{%
		Centre de Nanosciences et de Nanotechnologies (C2N), CNRS, Univ. Paris-Sud, Universit\'e Paris-Saclay, 10 bvd Thomas Gobert, 91120 Palaiseau, France
    }
	\affiliation{%
        The Institute for Nuclear Problems, Belarusian State University, 11 Bobruiskaya  Street, 220030, Minsk, Belarus
	}%
	\author{Fabrice Boust}%
	%\email{fabrice.boust@onera.fr}
	\affiliation{%
		SONDRA, CentraleSup\'elec, Universit\'e Paris-Saclay,
		F-91190, Gif-sur-Yvette, France
	}
	\affiliation{%
		DEMR, ONERA, Universit\'e Paris-Saclay, F-91123, Palaiseau, France
	}%
	
	%\altaffiliation

	\author{Jean-Luc Pelouard}
	\affiliation{%
		Centre de Nanosciences et de Nanotechnologies (C2N), CNRS, Univ. Paris-Sud, Universit\'e Paris-Saclay, 10 bvd Thomas Gobert, 91120 Palaiseau, France
    }

	\author{Fabrice Pardo}
		\affiliation{%
		Centre de Nanosciences et de Nanotechnologies (C2N), CNRS, Univ. Paris-Sud, Universit\'e Paris-Saclay, 10 bvd Thomas Gobert, 91120 Palaiseau, France
    }

	\author{Shah Nawaz Burokur}%
	%\email{sburokur@parisnanterre.fr}
	\affiliation{%
		LEME, UPL, Univ Paris Nanterre, F92410, Ville d'Avray, France
		%This line break forced with \textbackslash\textbackslash
	}%

\begin{abstract}
Recently, metamaterials-inspired diffraction gratings (or metagratings) have demonstrated unprecedented  efficiency in wavefront manipulation  by means of relatively simple structures. Conventional one-dimensional (1D) gratings have a profile modulation in one direction and a translation symmetry in the other.  In 1D metagratings, the translation invariant direction is engineered at a subwavelength scale what allows one to accurately control polarization line currents and, consequently, the scattering pattern. In bright contrast to metasurfaces, metagratings cannot be described by means of surface impedance densities (or local reflection and transmission coefficients). 
In this paper, we  present a simulation-based design approach to construct  metagratings in the ``unit cell by unit cell'' manner.
It represents an analog of the local periodic approximation (LPA) that has been used to design space modulated metasurfaces and allows one to overcome the limitations of straightforward numerical optimization and semi-analytical procedures that have been used up to date to design metagratings.
Electric and magnetic metagrating structures responding to respectively transverse electric (TE) and transverse magnetic (TM) incident plane-waves are presented to validate the proposed design approach. 
\end{abstract}

\maketitle

%\tableofcontents

\section{introduction}

In the last decade there has been tremendous interests in metasurfaces due to their amazing  capabilities in manipulating electromagnetic fields~\cite{kildishev2013planar,Glybovski2016,asadchy2018bianisotropic} and broad range of potential applications~\cite{Capasso2014_review_MSs,Yu2016_review,asadchy2018bianisotropic,capasso2018future}. 
Metasurfaces are  represented by dense distributions of engineered subwavelength scatterers on a surface being planar analogs of metamaterials. 
When the characteristics of a metasurface are spatially modulated, it can perform  wavefront transformations~\cite{Capasso_GeneralizedReflectionLaw,Pfeiffer2013}.
The local periodic approximation (LPA) plays a crucial role in designing such metasurfaces and  has been already used for long time~\cite{Pozar1997_LPA,Epstein2016_HMS_review}. Essentially, it serves to estimate scattering properties of a unit cell embedded in a nonuniform array. To that end, a unit cell is placed in the corresponding uniform array whose reflection and transmission coefficients are then attributed to the unit cell in a nonuniform array. Scattering parameters of a uniform array are usually calculated from full-wave numerical simulations. However, there are particularly simple cases (e.g. metallic patches) that can be treated analytically~\cite{Tretyakov_patches_2008,Tretyakov_MetaAnalyt_2018}.

\begin{figure}[tb]
\includegraphics[width=0.99\linewidth]{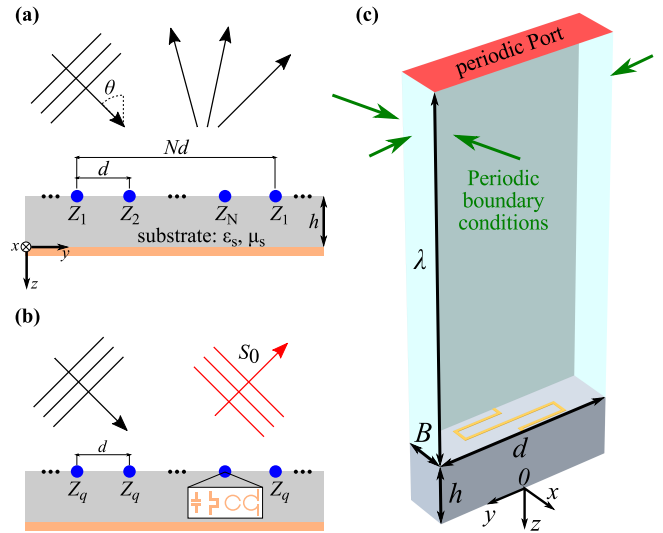}
\caption{\label{fig:1} (a) Schematic diagram of a metagrating: a periodic array of thin ``wires''  placed on a  dielectric substrate backed by a metal plate and  having relative permittivity $\E_s$, permeability $\mu_s$ and thickness $h$. The array is excited by a plane-wave incident at an angle $\theta$.
(b) Schematic diagram of a uniform array of ``wires'' characterized by the same impedance density $Z_q$. The inset represents the different meta-atoms composing a ``wire''. (c) Principal model used in numerical simulations to calculate the reflection coefficient from a uniform array of ``wires'' implemented with meta-atoms.}
\end{figure}

Recently, metamaterials-inspired diffraction gratings have demonstrated unprecedented  efficiency in manipulating  scattering patterns with relatively simple structures~\cite{Sell2017,Alu2017_metagrating,Epstein2017_metagrating,Popov2018,Epstein2018_exp_anrefr,Popov2018_perfect}. 
Reflecting configuration of a metagrating  represented by a one-dimensional periodic  array of   thin ``wires'' placed on top of a metal-backed dielectric substrate is illustrated  in Fig.~\ref{fig:1} (a).
Generally, each period of a metagrating consists of $N$ different ``wires''.
Noteworthy, the distance $d$ between adjacent  ``wires'' always remains of the order of operating wavelength $\lambda$, which does not allow one to introduce neither averaged surface impedances nor local reflection coefficient in contrast to metasurfaces.
Incident wave excites electric or magnetic polarization line currents in ``wires'' that can be controlled by judiciously adjusting the electromagnetic response of the ``wires''. Consequently, it becomes possible to manipulate diffraction orders.
It has been demonstrated that it is possible to achieve perfect nonspecular reflection and beam splitting with a single ``wire''  per period~\cite{Alu2017_metagrating,Epstein2017_metagrating,Epstein2018_exp_anrefr}. In a more general manner, it was shown that in order to perfectly control diffraction patterns  one needs two   degrees of freedom (represented by reactively loaded ``wires'') per each propagating diffraction order~\cite{Popov2018_perfect}. However, even having the number of  reactive ``wires'' per period \textit{equal} to the number of propagating diffraction orders enables to perform efficient multichannel reflection, as demonstrated in~\cite{Popov2018}.

Practically,  a ``wire'' is constituted from subwavelength meta-atoms arranged in a dense uniform 1D array . Geometrical parameters of meta-atoms determine electromagnetic response of a ``wire''.
Up to date metagratings have been designed either by performing 3D full-wave numerical optimization of a whole metagrating's period~\cite{Alu2017_metagrating} or semi-analytically~\cite{Alu2017_two_UCs,Epstein2017_metagrating,Epstein2018,Popov2018,Popov2018_perfect}. While the first approach can be very time consuming when it comes to designing metagratings having many ``wires'' per period, the second one allows one to consider only very simple meta-atoms such as printed capacitors~\cite{Epstein2017_metagrating} or dielectric cylinders~\cite{Alu2017_two_UCs}.
In this paper, we develop the local periodic approximation for designing metagratings  with the help of 3D full-wave numerical simulations. 
In comparison to a straightforward numerical optimization, it significantly reduces the time spent on the design of metagratings since within the LPA one deals with a single unit cell at a time [meaning that we consider a uniform array formed by a given unit cell as illustrated by Fig.~\ref{fig:1}(b)].
Contrary to simple analytical models, the LPA also allows one to deal with complex meta-atom's geometries and accurately account for interactions between adjacent ``wires''.

The outline of the paper is as follows. In the second section we describe a retrieval technique which is used to extract characteristics of a ``wire'' from scattering parameters of the corresponding uniform array. 
We consider ``wires'' possessing either electric or magnetic responses which allows one to deal with both TE and TM polarizations.
In the same section, we outline a  model that can be used in a full-wave simulation software to construct a look-up table connecting retrieved characteristics of ``wires'' with corresponding parameters of meta-atoms.
The third section is devoted to validation examples where we demonstrate metagratings' designs operating in microwave and infrared frequency domains.
In the fourth section we discuss remained challenges and conclude the paper.

\begin{figure*}[tb]
\includegraphics[width=0.99\linewidth]{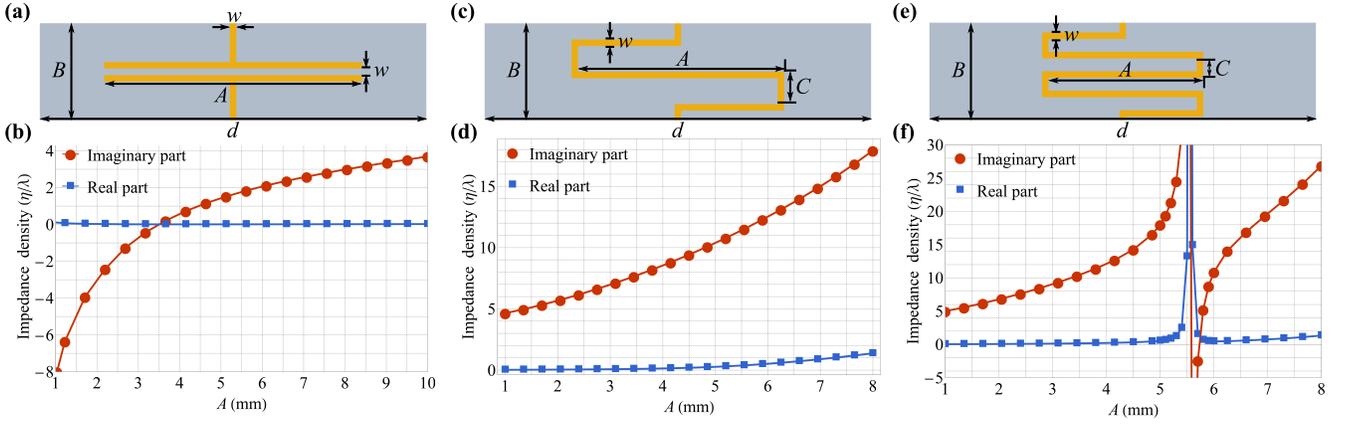}
\caption{\label{fig:2} Examples and characteristics of unit cells that can be used to compose ``wires'' in a metagrating. The top row represents the schematics of printed capacitor (a) and inductances (c) and (e). The bottom row demonstrates impedance densities found by means of the LPA as functions of geometrical parameters of unit cells and corresponding (from (b) to (f)), respectively, to ``wires'' built up from  printed capacitor shown in figure (a) and printed inductances in figures (c) and (e). Other parameters are fixed: dielectric substrate is F4BM220 of permittivity $\E_s=2.2(1-10^{-3}i)$ and thickness $h=5$ mm, $B=3.75$ mm, $d=15$ mm, $w=0.25$ mm, ($B/C=3$ in figure (d) and $B/C=5$ in figure (f)). Working frequency is set to $10$ GHz, corresponding to the vacuum wavelength of $30$ mm. Normally incident plane-wave is assumed where electric field is oriented along the $B$ dimension.}
\end{figure*}

\section{Local periodic approximation\\ for metagratings}

%For sake of simplicity 
We consider  reflecting  metagratings operating either under TE or TM incident wave polarization. 
Therefore, each ``wire''  composing  a metagrating can be characterized by a \textit{scalar} electric  impedance density $Z_q$ (or scalar magnetic admittance density $Y_q$ in  the case of a ``wire'' possessing magnetic response).
However, the approach can be readily generalized to transmitting  metagratings as it is discussed in the last section, polarization insensitive metagratings can be developed by designing  ``wires'' having both electric and magnetic responses and applying the method developed further for each polarization separately.
Previously, one has distinguished between load- and input-impedance densities~\cite{Epstein2017_metagrating,Popov2018,Popov2018_perfect} which we do not separate in the present study dealing only with impedance density as the principal characteristic of a ``wire''. 
To be more accurate, we assume that the impedance density represents the sum of the load-impedance density and reactive part of the input-impedance density.
According to the LPA, in order to find impedance density, a ``wire'' is placed in the corresponding uniform array (period $d<\lambda$) illuminated by a plane-wave incident at angle $\theta$ as illustrated by Fig.~\ref{fig:1} (b). 
We start by describing a way to retrieve  electric impedance $Z_q$ and magnetic admittance $Y_q$ densities from scattering parameters. 
Since $Z_q$  ($Y_q$) represents itself as a complex number and an electric (magnetic) line current radiates TE (TM) wave, it is sufficient to measure only the complex amplitude of the specularly reflected TE (TM) plane-wave.
Noteworthy,  time dependence $\exp[i\omega t]$ is assumed throughout the paper.

\subsection{Electric response, TE polarization}

An electric polarization line current $I$ (excited by TE plane-wave in a ``wire'' composing the uniform array) is linked to the complex amplitude $S_{0}^{TE}$ of the electric field of the specularly reflected wave via the following formula (see Appendix~\ref{app:A})
\begin{equation}
    \label{eq:I}
   I=-\frac{2d}{k\eta}\frac{(S_{0}^{TE}-R_0^{TE}e^{2i\B_0 h})\B_0}{(1+R_0^{TE})e^{i\B_0 h}}.
\end{equation}
Here, $k$ and $\eta$ are the wavenumber and characteristic impedance outside the substrate, $R_0^{TE}$ is the Fresnel's reflection coefficient from the metal-backed substrate of a TE plane-wave at incidence angle $\theta$ and $\B_0=k\cos(\theta)$.
Since all ``wires'' are assumed to be very thin, they are modeled as lines represented mathematically by the Dirac delta function $\delta(y,z)$.
Consequently, the interaction with the substrate and between adjacent ``wires'' can be taken into consideration analytically by means of the mutual-impedance  density $Z_m$  (see Appendix~\ref{app:A}). 
It allows one to obtain the characteristic of a ``wire'' itself and not of the array.
``Wire's'' electric impedance density   is found by means of Ohm's law leading to the following expression for $Z_q$ 
\begin{equation}
    \label{eq:Zq}
    Z_q=\frac{E_0}{I}-\frac{k\eta}{4}-Z_m,
\end{equation}
where $E_0=(1+R_0^{TE})\exp[i\B_0h]$ represents the value of the external electric field (incident wave plus its reflection from the metal-backed substrate) at the  the ``wire'' located at $y=0$ and $z=-h$.  The radiation resistance of a ``wire'' is equal to $k\eta/4$ being independent on its particular implementation as it follows from power conservation conditions~\cite{tretyakov2003analytical}. 
It is important to note that mutual-impedance density depends only on the period of the uniform array and parameters of the metal-backed substrate, but does not depend on the current $I$. It means that the impedance density given by Eq.~\eqref{eq:Zq} accurately represents the characteristic of the corresponding ``wire'' in a nonuniform array as long as the distance between adjacent ``wires'' and the metal-backed substrate remain the same as those of the uniform array.
However, we leave open the questions of  accuracy when considering ``wires'' built up from finite size meta-atoms as infinitely thin and estimating mutual interactions by means of analytically calculated mutual-impedance density.

\begin{figure}[tb]
\includegraphics[width=0.99\linewidth]{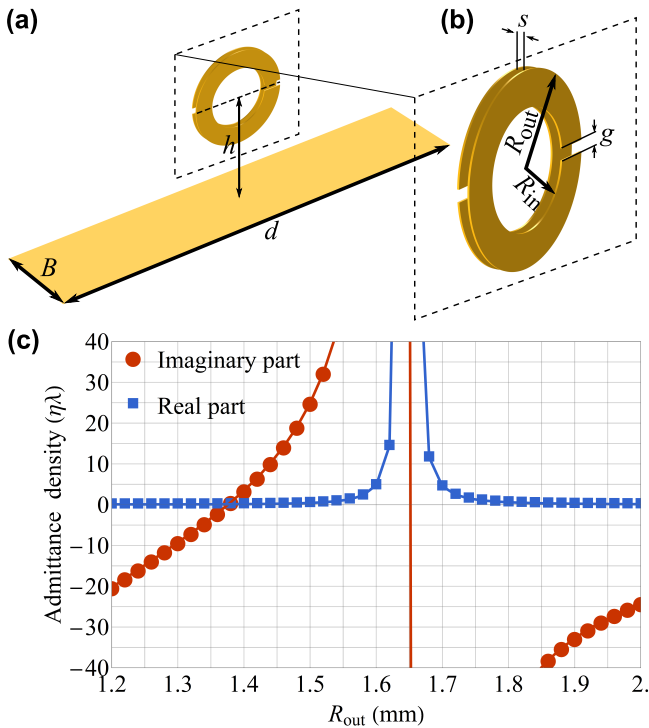}
\caption{\label{fig:3} (a) Schematic of a  unit cell based on two split ring resonators and having an inversion center. (b) Zoom view of the magnetic meta-atom with defined parameters. (c) Calculated by means of the LPA, admittance density of the ``wire'' built up from the magnetic meta-atoms versus the outer radius $R_\textup{out}$ of the split ring resonator. Other parameters are fixed: no substrate (air is considered as spacer), $h=3.75$ mm, $B=3.75$ mm, $d=15$ mm, $s=0.15$ mm, $g=0.20$ mm. Working frequency is set to $10$ GHz (vacuum wavelength of $30$ mm). Normally incident plane-wave is assumed where magnetic field is oriented along the $B$ dimension.}
\end{figure}

\subsection{Magnetic response, TM polarization}

The case of TM polarization and  ``wires'' possessing magnetic response   can be treated with the help of duality relations~\cite{felsen1994radiation}: $\textbf{E}\rightarrow\textbf{H}$, $\textbf{H}\rightarrow-\textbf{E}$, $I\rightarrow V$ and $\eta\rightarrow1/\eta$. 
Since the metal-backed dielectric substrate is not replaced by the corresponding dual equivalent, we have to additionally make the following substitution  $R_0^{TE}\rightarrow R_0^{TM}$.
Thus, from Eq.~\eqref{eq:I} one can arrive at the formula for retrieving the magnetic current $V$ from the complex amplitude of the magnetic field of the specularly reflected plane-wave (see Appendix~\ref{app:B})
\begin{equation}
    \label{eq:V}
   V=-\frac{2d\eta}{k}\frac{(S_{0}^{TM}-R_0^{TM}e^{2i\B_0 h})\B_0}{(1+R_0^{TM})e^{i\B_0 h}},
\end{equation}
where $R_0^{TM}$ is the Fresnel's reflection coefficient from the metal-backed substrate of a TM-polarized plane-wave at incidence angle $\theta$.
As previously, the interaction with the substrate and between adjacent ``wires'' can be taken into account by means of the mutual-admittance  density $Y_m$  calculated analytically (see Appendix~\ref{app:B}). Then, the magnetic admittance density $Y_q$ can be found as
\begin{equation}
    \label{eq:Yq}
    Y_q=\frac{H_0}{V}-\frac{k}{4\eta}-Y_m.
\end{equation}
Here $H_0=(1+R_0^{TM})\exp[i\B_0h]$ is the value of the external magnetic field at the ``wire'' located at $y=0$ and $z=-h$, $k/(4\eta)$ represents the radiation conductance. 

\subsection{Look-up table}

As it is stated above, in practice a ``wire'' would be implemented with subwavelength meta-atoms arranged in a line. 
The ultimate goal of the developing approach is to construct a look-up table linking geometrical parameters of meta-atoms with corresponding impedance (admittance) densities.
To that end, we harness  3D full-wave numerical simulations (in our examples  \uppercase{comsol multiphysics} is used). 
Although, Eqs.~\eqref{eq:I}--\eqref{eq:Yq} are obtained by modelling “wires” as infinitesimally thin, all practical features of unit cells (like dielectric and conduction losses, finite thickness of metallic traces, etc.) are taken into account by means of numerical simulations.
The geometry of the model consists of two principal parts: a tested unit cell (illustrated by a printed inductance on a metal-backed dielectric substrate in Fig.~\ref{fig:1} (c)) and air region.  Periodic boundary conditions are imposed on the side faces (as shown in Fig.~\ref{fig:1} (c)). The model is excited by a periodic port assigned to the face of the air region in opposite to the unit cell, as highlighted in red color in Fig.~\ref{fig:1} (c). The periodic Port creates a  plane-wave incident at angle $\theta$. It is important to take into account  $\theta$  since  meta-atoms are usually spatially dispersive (impedance density depends on the incidence angle, see for instance Ref.~\cite{Epstein2016_HMS_review}). The thickness of the air region equals operating vacuum wavelength $\lambda$ which is normally enough to eliminate higher order evanescent modes.
The periodic port is also used as a listening port to calculate the scattering parameter $S_{11}$. It is  related to the complex amplitude $S_0$ in Eq.~\eqref{eq:I} as $S_{11}=S_0^{TE}e^{-2i\B_0(h+\lambda)}$.
In the case of TM polarization $S_{11}=-S_0^{TM}e^{-2i\B_0(h+\lambda)}$.

\section{examples}

In order to validate the developed  approach, we first employ it to construct look-up tables that are further used to implement metagratings for controlling propagating diffraction orders. In order to find impedance densities necessary to establish a given diffraction pattern  we use the theoretical approach developed in Ref.~\cite{Popov2018}. 
In what follows, we focus on two frequency ranges: microwave (operating vacuum wavelength $30$ mm) and infrared (operating vacuum wavelength $4$ $\mu$m).

\subsection{Microwave frequency range}

\begin{figure}[tb]
\includegraphics[width=0.9\linewidth]{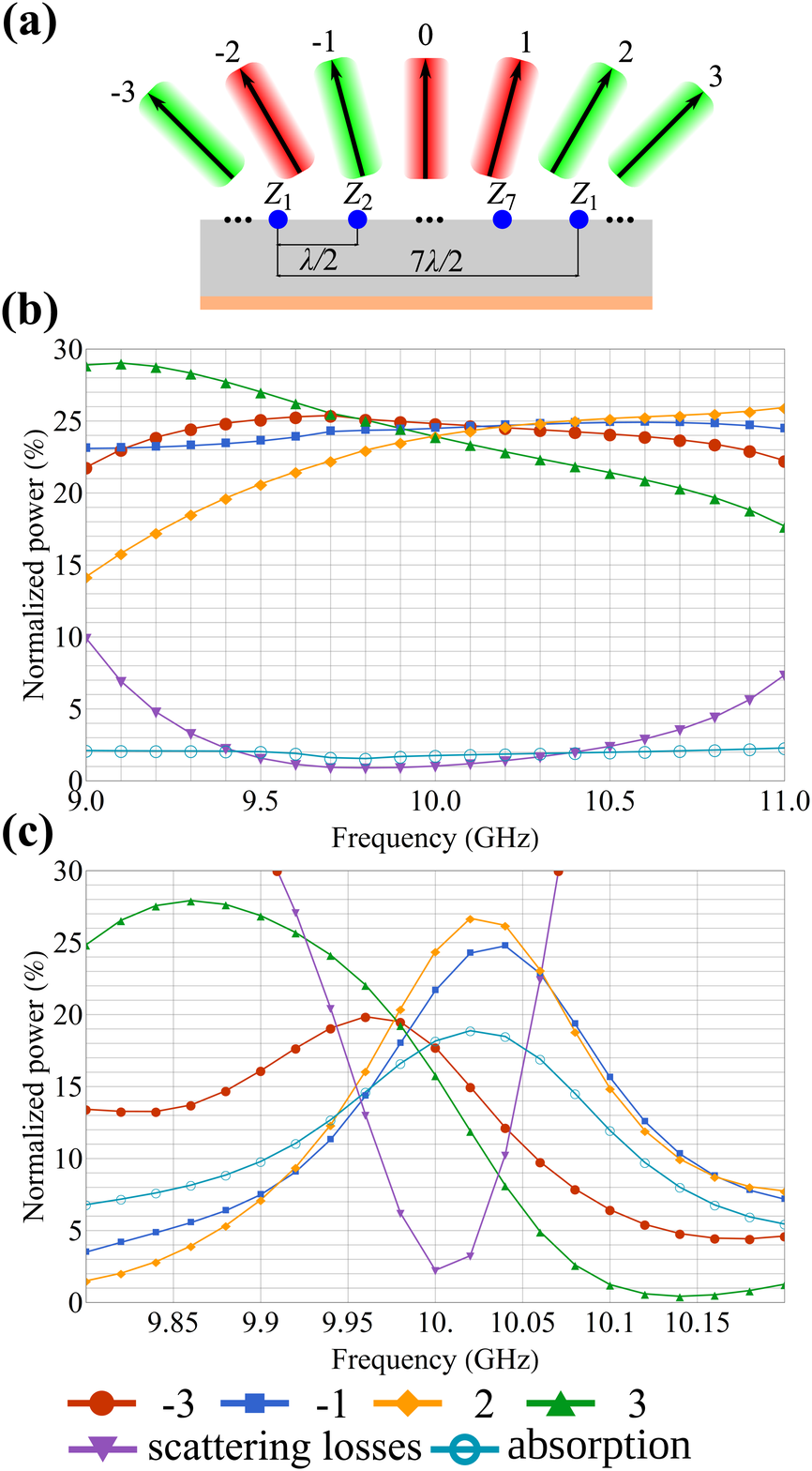}
\caption{\label{fig:4} (a) Schematic of a metagrating having period  $7\lambda/2$ ($\lambda$ is the operating vacuum wavelength) and exciting seven propagating diffraction orders under normally incident plane-wave. The red and green beams represent suppressed and equally excited orders, respectively. (b), (c) Simulated frequency response (normalized power scattered in propagating diffraction orders versus frequency) of the metagrating operating under (b) TE and (c) TM polarizations and establishing the diffraction pattern illustrated by figure (a). Both metagratings are designed to operate at $10$ GHz. Impedance and admittance densities as well as geometrical parameters of the ``wires'' composing the metagratings are given in Appendix C.}
\end{figure}

We start by considering simple  meta-atoms represented by printed capacitance and inductance (illustrated in  Fig.~\ref{fig:2}) that have already been used to implement metagratings at microwave frequencies by means of printed circuit board (PCB) technology~\cite{Epstein2018_exp_anrefr,Popov2018_perfect}. In the simulations, we take into account practical aspects of the design such as finite  thickness of the copper traces ($t_m=35$ $\mu$m) and dielectric losses introduced by the substrate (F4BM220 in our examples, $\E_s=2.2$ with loss tangent  $10^{-3}$).
Figure~\ref{fig:2} depicts the impedance densities calculated by means of the developed LPA at $10$ GHz (vacuum wavelength $\lambda$ is $30$ mm).
It is seen that having printed capacitance and inductance one is able to cover a broad range of impedance densities (imaginary part) that is normally enough to realize any diffraction pattern for TE polarization. 
It is important to note, that ``wires'' can exhibit significant resistive response (see Fig.~\ref{fig:2} (f) at the resonance) which should be kept in mind while designing metagratings.
When comparing the numerical results with analytical models used in Refs.~\cite{Epstein2017_metagrating,Popov2018_perfect}, one would see very good agreement for the imaginary part of the impedance density at small values of parameter $A$.
The resonance observed in Fig.~\ref{fig:2} (f) appears when one decreases the parameter $C$ and it cannot be found by simple analytical formula used in Ref.~\cite{Popov2018_perfect}.

In order to deal with TM polarization, we harness split ring resonators (SRRs) excited by the magnetic field and, thus, having effective magnetic response. Figure~\ref{fig:3} (a)  illustrates the schematics of the unit cell which at close look consists of two SRRs separated by a short distance as seen from Fig.~\ref{fig:3} (b).  The unit cell has an inversion center allowing to eliminate the bianisotropic response attributed to  single and double SRRs~\cite{Marques2002,SimovskiDSRR2004}.
In order to adjust the response of a ``wire'' represented by a 1D array of SRRs, we use  the outer radius $R_{\textup{out}}$ as a free parameter. 
The result of applying the LPA to find an equivalent admittance density is depicted in Fig.~\ref{fig:3} (c).
Due to the small separation distance between the two SRRs we are able to achieve strong magnetic response with the outer radius being of the order of $\lambda/20$.
As in the case of TE polarization discussed above we see that when approaching the resonance there is an increase of the real part of the ``wire's'' admittance density resulting in enhanced absorption.

In order to validate calculated impedance and admittance densities we develop  designs  of  metagratings based only on the data from Figs.~\ref{fig:2} and \ref{fig:3} to construct a certain diffraction pattern. 
Particularly, we demonstrate a splitting of normally incident plane-wave equally between four propagating diffraction orders (-3$^\textup{rd}$, -1$^\textup{st}$, +2$^\textup{nd}$ and +3$^\textup{rd}$) while suppressing the rest three (-2$^\textup{nd}$, 0$^\textup{th}$ and +1$^\textup{st}$), as illustrated by the schematic in Fig.~\ref{fig:4} (a). 
In accordance with Ref.~\cite{Popov2018}, to that end  we need the number of reactive ``wires'' per period equal to the number of propagating diffraction orders, i.e. seven. 
Although Ref.~\cite{Popov2018} deals only with electric line currents and TE polarization, it is straightforward to generalize the approach to magnetic currents and TM polarization (See Appendix~\ref{app:B}).
Figures~\ref{fig:4} (b) and (c) demonstrate the frequency response of electric and magnetic metagratings designed for $10$ GHz.
Overall, one can see that despite all practical limitations, the designed metagratings almost perfectly perform the desired splitting of the incident wave.
It is important to note that the response of the metagrating operating under TE polarization (Fig.~\ref{fig:4} (b)) shows a broader response than the one operating under TM polarization (Fig.~\ref{fig:4} (c)).
It is naturally explained by the resonant behavior of the considered SRR-based unit cell (see Fig.~\ref{fig:3} (b)).
Indeed, printed capacitance and inductance (illustrated, respectively, by Figs.~\ref{fig:2} (a) and (c)) used for the metagrating dealing with TE polarization do not exhibit resonances (see Figs.~\ref{fig:2} (b) and (d)).
The other  feature of the magnetic metagrating is the significant absorption (compared to the case of TE polarization) which, however, does not deteriorate the overall performance.

\begin{figure}[tb]
\includegraphics[width=0.9\linewidth]{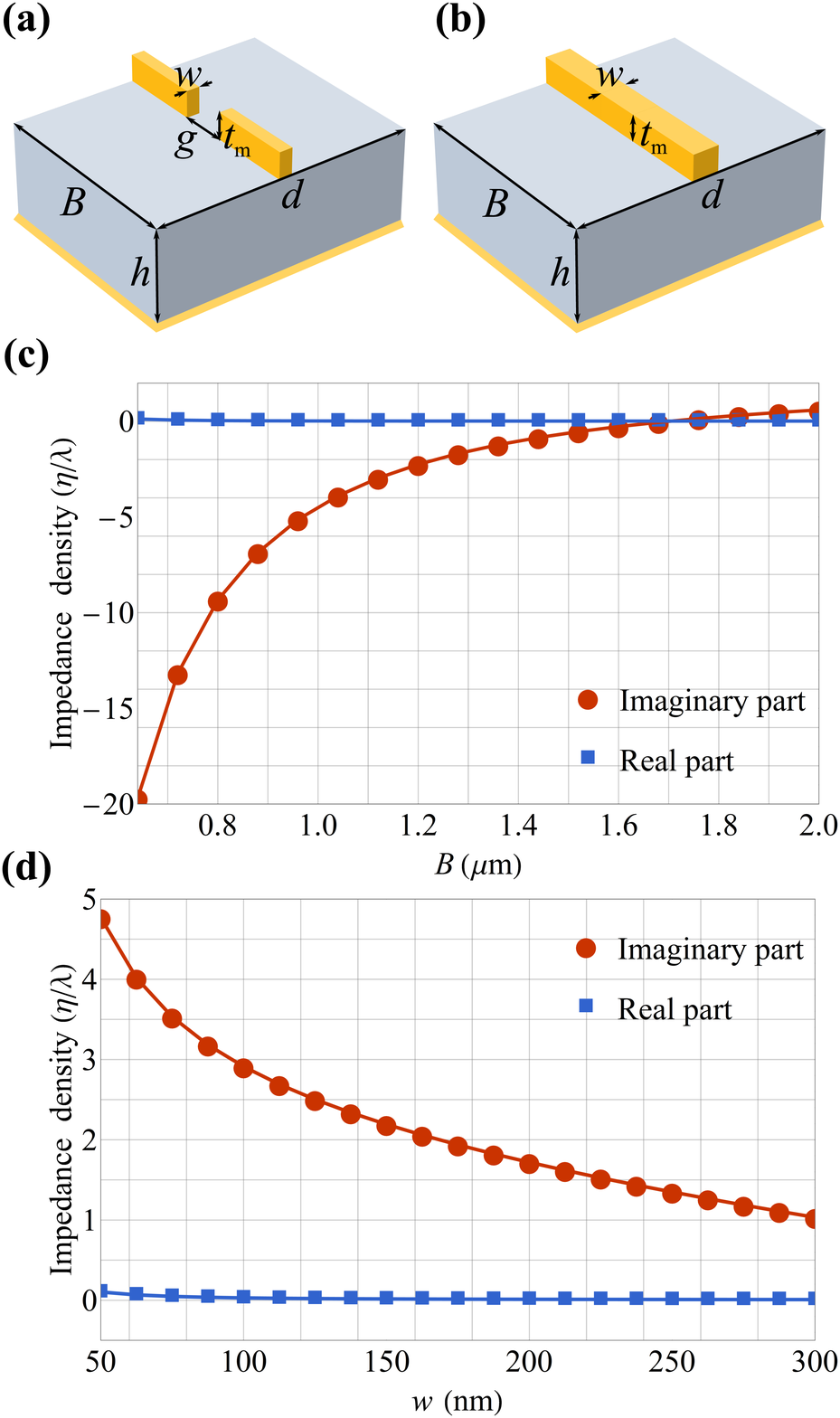}
\caption{\label{fig:5} (a), (b) Schematic diagrams of (a) two gold patches and (b) a gold ``wire'' exhibiting capacitive and inductive responses, respectively. The gold elements are placed on a silicon dioxide layer backed by a gold plating. (c), (d) Impedance densities calculated by means of the LPA as functions of geometrical parameters of the unit cells and corresponding, respectively, to ``wires'' built up from gold patches shown in figure (a) and gold wires in figure (b). Other parameters are fixed: silicon dioxide layer has permittivity $\E_s\approx1.93$ and thickness $h=700$ nm,  $d=2$ $\mu$m, $w=200$ nm, $t_m=200$ nm ($t_m=w$ in case of figure (d)), $g=350$ nm. Working frequency is set to $75$ THz (vacuum wavelength of $4$ $\mu$m). Normally incident plane-wave is assumed where electric field is oriented along the $B$ dimension.}
\end{figure}

\subsection{Infrared frequency range}

In this subsection we give an example of possible designs of unit cells that can be used as building blocks for metagratings operating at infrared frequencies.
In what follows, the operation frequency is set to $75$ THz corresponding to the vacuum wavelength of $4$ $\mu$m.
In order to implement capacitive and inductive unit cells for infrared domain we consider metallic (gold) patches and wires. Gold elements are placed on a dielectric substrate (silicon dioxide) backed with gold  as illustrated in Figs.~\ref{fig:5} (a) and (b).
The capacitive response is attributed to the gap between two patches. By changing the size of the gap or the size of patches one is able to adjust the capacitance. 
The inductance of a metallic wire is determined only by the cross section area.
Although the  design of the unit cells is relatively simple, it allows one to obtain impedance densities in a quite wide range of values as shown in Figs.~\ref{fig:5} (c) and (d).
Interestingly, the real part of the impedance density remains very small due to nonresonant response of the unit cells even though we take into account absorption in gold and silicon dioxide.

It is worthwhile to note that a single straight metallic wire may not be enough in case  strong inductive response is necessary (large positive imaginary part of the impedance density).
Cross section area is usually restricted by fabrication tolerances that does not allow one to infinitely reduce it. 
Meandering can be a solution (see Figs.~\ref{fig:2} (c) and (e)) though it might complicate the fabrication.

%Overall, extracted impedance densities slightly depend on the permittivity of gold since the skin depth at 4 $\mu$m constitutes approximately $25$ nm what is less than the geometrical dimensions of the elements.

\begin{figure}[tb]
\includegraphics[width=0.9\linewidth]{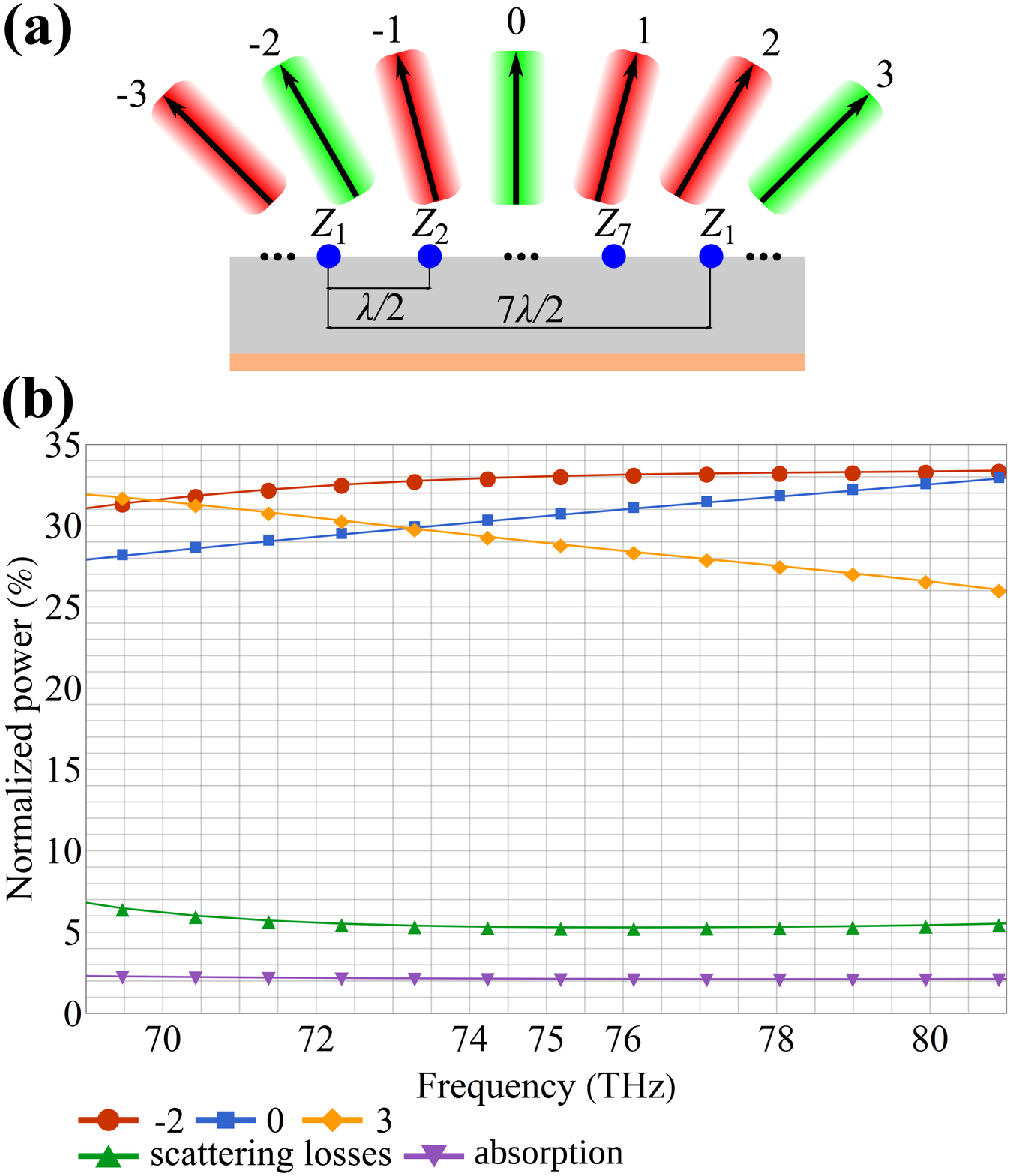}
\caption{\label{fig:6} (a) Schematic diagram of a metagrating having period  $7\lambda/2$ ($\lambda$ is the operating vacuum wavelength) and exciting seven propagating diffraction orders under normally incident plane-wave. The red and green beams represent suppressed and equally excited orders, respectively. (b) Simulated frequency response (normalized power scattered in different propagating diffraction orders versus frequency) of the metagratings operating under  TE polarization and establishing the diffraction pattern illustrated by figure (a). The metagrating is designed for $75$ THz operating frequency. Impedance densities as well as geometrical parameters of the ``wires'' composing the metagrating are given in Appendix C.}
\end{figure}

To validate calculated impedance densities we demonstrate a metagrating equally splitting a normally incident plane-wave between three propagating diffraction orders (-2$^\textup{nd}$, 0$^\textup{th}$ and +3$^\textup{rd}$). The rest four propagating orders are cancelled, as presented by the schematic in Fig.~\ref{fig:6} (a).
Simulated frequency response of the metagrating is demonstrated in Fig.~\ref{fig:6} (b).
The incident wave equally (with an accuracy of a few percentages) excites chosen diffraction orders.
One can see that the scattering in the -3$^\textup{rd}$, -1$^\textup{st}$, +1$^\textup{st}$ and +2$^\textup{nd}$ orders remains suppressed in a wide frequency range due to the nonresonant nature of the unit cells and that absorption constitutes only $2\%$ of total power.
The $5\%$ level of scattering losses is rather a drawback of the approach presented in Ref.~\cite{Popov2018}, but can still be reduced as proposed by the design methodology of Ref.~\cite{Popov2018_perfect}.

To conclude this subsection, let us make a few practical comments.
In the numerical simulations, dielectric permittivities of gold and silicon dioxide  were taken from Refs.~\cite{Babar2015} and \cite{Kischkat2012}, respectively.
Silicon dioxide was chosen as a substrate due to low losses and low refractive index at $4$ $\mu$m. 
It allows one using a thick substrate while avoiding excitation of waveguide modes. On the other hand, if the dielectric substrate is thin, it may not be possible to model ``wires'' as having  infinitely small cross section area and accurately account for the interaction with the substrate.
Another practical feature is that between gold parts and a silicon dioxide substrate there is usually a thin chromium (or titanium) adhesion layer of a few nm thickness.
Although all practical aspects should be taken into account while designing an experimental sample, we do not expect that the influence of such a thin intermediate layer can lead to a qualitative change and, therefore,  did not consider it. 

\section{discussion and conclusion}

In this work we have presented a simulation-based design approach to construct  metagratings in the ``unit cell by unit cell'' manner.
It represents an analog of the local periodic approximation that has been used to design space modulated metasurfaces and in comparison to a brute force numerical optimization (that deals straight with a whole supercell of a metagrating) can considerably reduce the time spent on the design.
Indeed, assume that each unit cell constituting a $N$-cells period of a metagrating has only one parameter to be adjusted and this parameter takes in total $P$ different values during a parametric sweep. 
Then,  one would have to perform $P^N$ full-wave simulations in order to  cover all possible combinations of parameters of unit cells and find optimal configuration of a metagrating's supercell [assuming no ``smart'' algorithms (like genetic) are harnessed]. Clearly, this number strongly depends on the number of unit cells in a supercell $N$.
Meanwhile, the LPA deals with one unit cell at a time what makes the number of required full-wave simulations can be as small as $P$ and it does not depend on the number $N$.
We have validated the developed approach via 3D full-wave numerical simulations by demonstrating designs of metagratings controlling scattering patterns at microwave and infrared frequencies for both TE and TM polarizations.
The both electric metagratings (TE polarization) required two different types of unit cells (capacitive and inductive) and overall we have performed only $2P$ simulations, while the magnetic metagrating required one type of a unit cell and, thus, only $P$ simulations.
Finally, simple and accurate analytical model describing metagratings allows one to take into account the impact of the metal-backed dielectric substrate and the interaction between ``wires'' composing the metagratings.
It makes the local periodic approximation not only fast but also  a  rigorous approach  to design metagratings represented by nonuniform arrays of  ``wires'' (in bright contrast to metasurfaces). 

In this work, we have considered a reflecting configuration of a 1D metagrating but the developed design approach can be generalized to deal also with transmitting and 2D metagratings.
Up to date there are two methods to control transmission with metagratings: either by means of an asymmetric three-layer array~\cite{epstein2018anrefr} of electric-only ``wires'' or a single-layer array of bianisotropic particles~\cite{fan2018perfect}.
In order to deal with 2D metagratings one would have to consider a 2D periodic array of point scatterers instead of a 1D array of line currents.
In light of recently demonstrated acoustic metagratings~\cite{Packo2018,Torrent2018} we also expect that the LPA can be adjusted to design acoustic wavefront manipulation devices.

To conclude, wavefront control  by means of metagratings is not only an innovative and efficient way to realize the transformation between incident and reflected and/or transmitted waves, but it also allows one to  further reduce the  fabrication complexity  while significantly improving the performance and functionality  of metagratings-based  integrated optics components  and reconfigurable antennas in microwave communication systems. 
In this respect, the developed  approach can be an essential tool when designing complex passive or active metagratings across all electromagnetic spectrum.

\section*{acknowledgement}

M.Y., J-L.P. and F.P. acknowledge financial support from ANR (grant mEtaNiZo).

% V.P. and M.Y. would like to thank Jean-Luc Pelouard (CNRS) and Fabrice Pardo (CNRS) for valuable and encouraging discussions on implementation of metagratings at infrared frequencies.

\appendix

\section{Retrieval of electric current and impedance density from specular reflection}\label{app:A}

In this section, we derive Eqs.~\eqref{eq:I} and \eqref{eq:Zq}. To that end we use analytical formulas for the electric field scattered by a $N$-cells metagrating given in Ref.~\cite{Popov2018}.
The system is excited by an incident plane  wave with the electric field $e^{-i k \sin \theta y - i k \cos \theta z + i \omega t}$ being along the $x$-direction.
Since in the LPA we simulate only a single ``wire'' per period, the electric field of the wave reflected from the periodic array depicted in Fig.~\ref{fig:1} (b) can be represented via plane-wave expansion as follows
\begin{eqnarray}\label{eq:A:Ex}
  &&E_x(y,z)=-I\frac{k \eta }{2d}\sum_{m=-\infty}^{+\infty}\frac{(1+R^{TE}_m)}{\B_m}e^{-i\xi_m y+i\B_m(z+h)}\nonumber\\
  &&+R_0^{TE} e^{-i\xi_0 y+i\B_0(z+2h)},
\end{eqnarray}
where $\xi_m=k\sin\theta+2\pi m/d$ and $\B_m=\sqrt{k^2-\xi_m^2}$ are, respectively, tangential and normal components of the wave vector of the $m^\textup{th}$ diffraction order. The Fresnel's reflection coefficient from the  substrate backed by perfect electric conductor (PEC) is given by the following formula
\begin{equation}\label{eq:A:RTE}
    R_m^{TE}=\frac{i\gamma_m^{TE}\tan[\B_m^sh]-1}{i\gamma_m^{TE}\tan[\B_m^sh]+1},\quad \gamma_m^{TE}=\frac{k_s\eta_s\B_m}{k\eta \B_m^s},
\end{equation}
where $\eta=\sqrt{\mu/\E}$ and $\eta_s=\sqrt{\mu_s/\E_s}$.
When a metal backing the dielectric substrate cannot be modeled as PEC, one has to correspondingly modify the reflection coefficient.
The period $d$ and incidence angle $\theta$ are such that the  only propagating diffraction order corresponds to the specular reflection ($m=0$).
Then, the amplitude $S_0^{TE}$ of the zeroth diffraction order can be found from Eq.~\eqref{eq:A:Ex} to be
\begin{equation}\label{eq:A:S0TE}
    S_0^{TE}=R_0^{TE} e^{2i\B_0h}-I\frac{k \eta }{2d}\frac{1+R^{TE}_0}{\B_0}e^{i\B_0h}.
\end{equation}
This formula is then used to express the current $I$ leading to Eq.~\eqref{eq:I}.

In order to calculate the impedance density of a ``wire'' one needs to find the ratio between the  total electric field $E_{(loc)}$ at the position of a  ``wire'' and the current $I$ in the ``wire'' (and then subtract the radiation resistance). $E_{(loc)}$  can be represented by the sum of the external electric field $E_0$, the electric field created by all other line currents and the field resulted from the reflection from the metal-backed substrate. 
The last two can be united and expressed as $-Z_mI$ with the mutual-impedance density $Z_m$ given by the following formula
\begin{eqnarray}\label{eq:A:Zm}
  &&Z_m=\frac{k\eta}{2}\sum_{n=1}^{+\infty} \cos[k\sin[\theta]nL]H_0^{(2)}[knd]\nonumber\\&&+\frac{k\eta}{2d}\sum_{m=-\infty}^{+\infty}\frac{R_m^{TE}}{\B_m}.
\end{eqnarray}
 Thus one arrives at the Eq.~\eqref{eq:Zq}.
 
\section{Magnetic metagratings}\label{app:B}

A reflecting metagrating possessing magnetic-only response is modeled as a periodic array of magnetic line currents placed on top of a metal-backed dielectric substrate. 
In the first two subsections we derive  analytical formulas (following Ref.~\cite{Popov2018}) for the field radiated by such a system and describe the way towards control of diffraction patterns with magnetic metagratings.
In the last subsection we derive Eqs.~\eqref{eq:V} and \eqref{eq:Yq}.

\subsection{Radiation of an array of magnetic line currents}
A single magnetic line current $\bf M(\bf r)=V\delta(y,z)\bf x_0$  radiates a TM-polarized wave with the magnetic field being along the $x$-direction (see Ref.~\cite{felsen1994radiation})
\begin{equation}\label{eq:MLC_x}
H_x(y,z)=-\frac{k}{4\eta}V H_0^{(2)}\left[k\sqrt{y^2+z^2}\right],
\end{equation}
where $H_0^{(2)}\left[k\sqrt{y^2+z^2}\right]$ is the Hankel function of the second kind and zeroth order.
Consequently,  the magnetic field radiated by a periodic array of $N$ magnetic line currents per period
$\bf M_{nq}(\bf r)=V_{q}\exp[-ik\sin[\theta]nL]\delta(y-y_{nq},z) \bf x_0$ (see Fig.~\ref{fig:1} (a)) is given by the series of Hankel functions
\begin{eqnarray}
\label{eq:MLCA_x0}
&&H_x(y,z)=-\frac{k}{4\eta}\sum_{q=1}^{N}\sum_{n=-\infty}^{+\infty} V_q e^{-ik\sin[\theta]nL}\nonumber\\
&&H_0^{(2)}[k\sqrt{(y-n L-(q-1)d)^2+z^2}],
\end{eqnarray}
where $y_{nq}= nL+(q-1)d$, $n$ and $q$ take integer values from $-\infty$ to $+\infty$ and from $1$ to $N$, respectively.
Since the array is periodic the field given by Eq.~\eqref{eq:MLCA_x0} can be expressed via the series of plane-waves by means of the Poisson's formula (see Ref.~\cite{Popov2018}) as follows
\begin{equation}\label{eq:MLCA_x}
H_x(y,z)=-\frac{k}{2\eta L}\sum_{m=-\infty}^{+\infty}\frac{\rho_m^{(V)}}{\B_m}e^{-i\xi_m y}e^{-i\B_m |z|},
\end{equation}
where $\rho_m^{(V)}=\sum_{q=1}^NV_q\exp[i\xi_m(q-1)d]$.

The effect of the metal-backed dielectric substrate on the field radiated by the array $\bf M_{nq}(\bf r)=V_{q}\exp[-ik\sin[\theta]nL]\delta(y-y_{nq},z+h) \bf x_0$ can be derived  following  Ref.~\cite{Epstein2018}. 
After some algebra, one would arrive at the following expressions for the magnetic field profile  outside the substrate ($z<-h$) 
\begin{eqnarray}\label{eq:MLC_TE_sub}
&&H_x(y,z<-h)=\nonumber\\
&&-\frac{k}{2\eta L}\sum_{m=-\infty}^{+\infty}\frac{\rho^{(V)}_m(1+R_m^{TM})}{\B_m}e^{-i\xi_my+i\B_m(z+h)}.
\end{eqnarray}
Here $R_m^{TM}$ is the Fresnel's reflection coefficient of a plane-wave (having tangential component of wave vector equal to $\xi_m$) from the metal-backed dielectric substrate
\begin{equation}
R_m^{TM}=-\frac{i\gamma_m^{TM}\tan[\B_m^sh]-1}{i\gamma_m^{TM}\tan[\B_m^sh]+1},\quad \gamma_m^{TM}=\frac{k\eta_s\B_m^s}{k_s\eta \B_m}.
\end{equation}

%%%%%%%%%%%%%%%%%%%%%%%%%%%%%%%%%%%%%%%%%%%%%%%%%%%%%%%%%%%%%%%%%%%%%%%%%%%%%%%%%%%%%%%%

\subsection{Controlling diffraction patterns with magnetic metagratings}

\begin{table*}[tb]
\resizebox{0.99\textwidth}{!}{%
\begin{tabular}{|c|c|c|c|c|c|c|c|} 
 \hline
Impedance density ($\eta/\lambda$) & $Z_1$ & $Z_2$ & $Z_3$ & $Z_4$ & $Z_5$ & $Z_6$ & $Z_7$\\ 
 \hline
Microwaves: TE & $-i4.87$ & $-i1.16$ & $i6.88$ & $i17.16$ & $-i0.33$ & $i3.97$&$-i0.81$\\  
 \hline
 Infrared: TE &  $-i0.30$ & $-i3.06$ & $-i3.20$ & $i3.61$ & $-i1.63$ & $-i1.04$&$-i11.59$\\ 
 \hline
 \hline
 Admittance density ($\eta\lambda$) & $Y_1$ & $Y_2$ & $Y_3$ & $Y_4$ & $Y_5$ & $Y_6$ & $Y_7$\\ 
 \hline
Microwaves: TM &  $-i3.61$ & $i4.42$ & $i0.89$ & $i3.05$ & $i0.31$ & $i0.71$&$i4.51$\\ 
 \hline
 \hline
Geometrical parameters (mm) & $A_1$ & $A_2$ & $A_3$ & $A_4$ & $A_5$ & $A_6$ & $A_7$\\ \hline
Microwaves: TE & $1.48$ (cap. uc) & $2.74$ (cap. uc) & $3.08$ (ind. uc) & $7.88$ (ind. uc) & $3.26$ (cap. uc) & $10.33$ (cap. uc) &$2.94$ (cap. uc)\\  
 \hline
 \hline
Geometrical  parameters (mm) & $R_{out,1}$ & $R_{out,2}$ & $R_{out,3}$ & $R_{out,4}$ & $R_{out,5}$ & $R_{out,6}$ & $R_{out,7}$\\ \hline
 Microwaves: TM & $1.348$ & $1.4065$ & $1.382$ & $1.397$ & $1.378$ & $1.381$&$1.4071$\\  
 \hline
 \hline
Geometrical parameters (nm) & $w_1,\; B_1, \; g_1$ & $w_2,\; B_2, \; g_2$ & $w_3,\; B_3, \; g_3$ & $w_4, t_{m,4}=w_4$ & $w_5,\; B_5, \; g_5$ & $w_6,\; B_6, \; g_6$ & $w_7,\; B_7, \; g_7$ \\ 
\hline
 Infrared: TE &  $175,\; 1600,\; 350$ & $59,\; 800,\; 100$ & $53,\;800,\;100$ & $73$ (ind. uc) & $157,\;800,\;100$ & $260,\;800,\;100$&$137,\;800,\;350$\\ 
 \hline
 \end{tabular}%
}
\caption{\label{tab:1}Parameters of the metagratings demonstrated as examples in  Section III. The indexes correspond to the numbered ``wires'' in Figs.~\ref{fig:4} (a) and \ref{fig:6} (a). The rest parameters are fixed and given in the captions to Figs.~\ref{fig:2}, \ref{fig:3} and \ref{fig:5}. Where it is necessary the type of a used unit cell (uc) is specified in the brackets.
In the example demonstrated by Fig.\ref{fig:4}(b), inductive unit cells are represented only by meanders with $B/C=3$, as illustrated in Fig.~\ref{fig:2}(c).
}
\label{tab:1}
\end{table*}

When a magnetic  metagrating is illuminated by a TM-polarized plane-wave incident at angle $\theta$, the scattered magnetic field can be represented as a superposition of plane-waves
$
\sum_{m=-\infty}^{+\infty}S_m^{TM}e^{-i\xi_my+i\B_mz}
$.
Amplitudes of the plane-waves are found analytically   (see previous subsection) and can be expressed as follows
\begin{equation}\label{eq:Am}
S_m^{TM}=-\frac{k}{2\eta L}\frac{
(1+R_m^{TM})e^{i\B_mh}}{\B_m}\rho_m^{(V)}+\delta_{m0}R_0^{TM}e^{2i\B_0h}
\end{equation}
where $\delta_{m0}$ is the Kronecker delta representing the reflection of the incident wave from the metal-backed substrate.
As in case of TE polarization described in Ref.~\cite{Popov2018}, Eq.~\eqref{eq:Am} demonstrates that  magnetic line currents  contribute to the scattered plane-waves via the parameter $\rho_m^{(V)}$. 
Magnetic line currents $V_q$ can be used to control plane-waves scattered in the far-field ($|\xi_m|<k$).
Since in practice it is more straightforward to deal with passive structures and also the excitation of magnetic line currents can be of particular difficulty, we are interested in the case when magnetic line currents are polarization currents excited by the incident plane wave in thin ``wires'' characterized by admittance densities  $Y_q$.
Then, necessary currents $V_q$ can be obtained by choosing admittance densities $Y_q$ from the following equation \begin{equation}\label{eq:Y}
\left(Y_q+\frac{k }{4\eta}\right)V_q=H_q-\sum_{p=1}^N Y_{qp}^{(m)}V_p.
\end{equation}
The right-hand side of Eq.~\eqref{eq:Y} represents the total magnetic  field at the location of the $q^\textup{th}$ ``wire'', where $H_q=(1+R_0^{TM})\exp[i\B_0h-i\xi_0 (q-1)d]$ is the external magnetic field, $Y_{qp}^{(m)}$ are the mutual-impedance densities which account for the interaction with the substrate and adjacent ``wires''. 
The magnetic field created by $q^\textup{th}$ line current from all periods (except the zeroth one) is given by the following series
\begin{equation}\label{eq:Y1}
-\frac{k}{2\eta}V_q\sum_{n=1}^{+\infty} \cos[k\sin[\theta]nL]H_0^{(2)}[kn L].    
\end{equation}
The magnetic field created by all other line currents can be accounted for as follows 
\begin{equation}\label{eq:Y2}
-\frac{k}{4\eta}\sum_{p=1, p\neq q}^{N}V_p\sum_{n=-\infty}^{+\infty}  e^{-ik\sin[\theta]nL} H_0^{(2)}[k|(q-p)d-n L|].
\end{equation}
Eventually, the waves reflected from the metal-backed substrate create the following magnetic field at the location of the $q^\textup{th}$ ``wire'' (zeroth period)
\begin{equation}\label{eq:Series_R}
-\frac{k}{2\eta L}\sum_{p=1}^{N}V_p\sum_{m=-\infty}^{+\infty} e^{i\xi_m (p-q)d}\frac{R_m^{TM}}{\B_m}.    
\end{equation}
In contrast  to the case of TE polarization discussed in Refs.~\cite{Epstein2018,Popov2018}, the series in Eq.~\eqref{eq:Series_R} does not converge. 
Indeed,  $R^{TM}_m$ has $(\E_s-1)/(\E_s+1)$ as limit when $m$ goes to infinity and $\B_m\sim-im$ for large $m$.
Meanwhile, it is well known that the harmonic series is divergent.
However, the divergence in Eq.~\eqref{eq:Series_R} is artificial and was brought when using the Poisson's formula (see Eq.~\eqref{eq:MLCA_x}).
Thus, we can avoid divergence by using the Poisson's formula backwards.
To that end, we perform the following transformation of the series in Eq.~\eqref{eq:Series_R}

\begin{widetext}
\begin{equation}\label{eq:Series_R1}
-\frac{k}{2\eta L}\sum_{m=-\infty}^{+\infty} e^{i\xi_m (p-q)d}\frac{R_m^{TM}}{\B_m}=-\frac{k}{2\eta L}\sum_{m=-\infty}^{+\infty} e^{i\xi_m (p-q)d}\frac{1}{\B_m}\left(R_m^{TM}-\frac{\E_s-1}{\E_s+1}\right)-\frac{k}{2\eta L}\frac{\E_s-1}{\E_s+1}\sum_{m=-\infty}^{+\infty} e^{i\xi_m (p-q)d}\frac{1}{\B_m}. 
\end{equation}
The first series on the right hand side of Eq.~\eqref{eq:Series_R1} is now converging while the second one contains the singularity and should be transformed by means of the Poisson's formula in the following way
\begin{eqnarray}
\label{eq:Series_R2}
&&-\frac{k}{2\eta L}\frac{\E_s-1}{\E_s+1}\sum_{m=-\infty}^{+\infty} e^{i\xi_m (p-q)d}\frac{1}{\B_m}=-\frac{k}{4\eta}\frac{\E_s-1}{\E_s+1}\sum_{n=-\infty}^{+\infty} e^{-ik\sin[\theta]nL} H_0^{(2)}[k|(q-p)d-n L|],\quad q\neq p, \nonumber\\   
&&-\frac{k}{2\eta L}\frac{\E_s-1}{\E_s+1}\sum_{m=-\infty}^{+\infty}\frac{1}{\B_m}=-\frac{k}{2\eta}\frac{\E_s-1}{\E_s+1}\sum_{n=1}^{+\infty} e^{-ik\sin[\theta]nL} H_0^{(2)}[k n L]-\frac{k}{4\eta}\frac{\E_s-1}{\E_s+1},\quad q=p.   
\end{eqnarray}
Summarizing Eqs.~\eqref{eq:Y1}, \eqref{eq:Y2}, \eqref{eq:Series_R} and \eqref{eq:Series_R2} one arrives at the explicit expression for the mutual-admittance density
\begin{eqnarray}\label{eq:Ym_full}
&&Y_{qp}^{(m)}=\left(1+\frac{\E_s-1}{\E_s+1}\right)\frac{k}{4\eta}\sum_{n=-\infty}^{+\infty}e^{-ik\sin[\theta]nL} H_0^{(2)}[k|(q-p)d-n L|] +\frac{k}{2\eta L}\sum_{m=-\infty}^{+\infty} e^{i\xi_m (p-q)d}\frac{1}{\B_m}\left(R_m^{TM}-\frac{\E_s-1}{\E_s+1}\right),\quad q\neq p,\nonumber\\
&&Y_{qq}^{(m)}=\frac{\E_s-1}{\E_s+1}\frac{k}{4\eta}+\left(1+\frac{\E_s-1}{\E_s+1}\right)\frac{k}{2 \eta}\sum_{n=1}^{+\infty} \cos[k\sin[\theta]nL]H_0^{(2)}[knL]+\frac{k}{2 \eta L}\sum_{m=-\infty}^{+\infty}\frac{1}{\B_m}\left(R_m^{TM}-\frac{\E_s-1}{\E_s+1}\right).
\end{eqnarray}
\end{widetext}

%%%%%%%%%%%%%%%%%%%%%%%%%%%%%%%%%%%%%%%%%%%%%%%%%%%%%%%%%%%%%%%%%%%%%

It is worth to note that from the mathematical point of view, it is not strictly correct to represent a diverging series as a sum of two terms as it is done in Eq.~\eqref{eq:Series_R1}. 
However, when comparing estimations of the mutual-admittance densities given by Eq.~\eqref{eq:Ym_full} with the ones obtained from 2D full-wave numerical simulations (see Ref.~\cite{Popov2018_perfect}), we observed a good agreement between the two in case of low permittivity substrates (e.g., $\E_s = 2.2$). A more rigorous treatment is necessary for dealing analytically with high permittivity substrates.

\subsection{Retrieval of magnetic current and admittance density from specular reflection}

In this subsection, we derive Eqs.~\eqref{eq:V} and \eqref{eq:Yq} of the main text following the same order as in Appendix~\ref{app:A}. The magnetic field of the wave reflected from the uniform array of magnetic line currents illustrated in Fig.~\ref{fig:1} (b) can be represented via the plane-wave expansion as follows
\begin{eqnarray}
  &&H_x(y,z)=-\frac{k}{2\eta d}V\sum_{m=-\infty}^{+\infty}\frac{(1+R^{TM}_m)}{\B_m}e^{-i\xi_m y+i\B_m(z+h)}\nonumber\\
  &&+R_0^{TE} e^{-i\xi_0 y+i\B_0(z+2h)}.
\end{eqnarray}
The amplitude of specularly reflected plane-wave (the only propagating diffraction order, $d<\lambda$) is then found to be (see also Eq.~\eqref{eq:Am})
\begin{equation}\label{eq:S0_TM}
S_0^{TM}=-\frac{k}{2d\eta}\frac{
(1+R_0^{TM})e^{i\B_0h}}{\B_0}V+R_0^{TM}e^{2i\B_0h}.
\end{equation}
From this equation it is straightforward to arrive at Eq.~\eqref{eq:V}.

The other equation (Eq.~\eqref{eq:Yq}) is readily obtained from Eq.~\eqref{eq:Y} when $N$ equals $1$. The mutual-admittance density in this case is given by the following expression (see Eq.~\eqref{eq:Ym_full})
\begin{eqnarray}
  &&Y_m=\left(1+\frac{\E_s-1}{\E_s+1}\right)\frac{k}{2\eta}\sum_{n=1}^{+\infty} \cos[k\sin[\theta]nL]H_0^{(2)}[knd]\nonumber\\&&+\frac{k}{2d\eta}\sum_{m=-\infty}^{+\infty}\frac{1}{\B_m}\left(R_m^{TM}-\frac{\E_s-1}{\E_s+1}\right)+\frac{\E_s-1}{\E_s+1}\frac{k}{4\eta}.
\end{eqnarray}
  
\section{Parameters of metagratings shown in examples}

Table~\ref{tab:1} provides impedance (admittance) densities as well as geometrical parameters of the ``wires'' composing the metagratings demonstrated as examples in  Section III.

\bibliography{bib}

\end{document}